\documentclass{optica-article}

\journal{opticajournal} 

\articletype{Research Article}

\usepackage{lineno}

\begin{document}

\title{Crossover from exciton-polariton condensation to photon lasing in an optical trap}

\author{M. Pieczarka,\authormark{1,*} D. Biega\'{n}ska,\authormark{1} C. Schneider,\authormark{2}, S. H\"{o}fling,\authormark{3} S. Klembt,\authormark{3} G. S\k{e}k,\authormark{1} and M. Syperek\authormark{1}}

\address{\authormark{1}Department of Experimental Physics, Faculty of Fundamental Problems of Technology, Wroc\l aw University of Science and Technology, Wybrze\.{z}e Wyspia\'{n}skiego 27, 50-370 Woc\l aw, Poland\\
\authormark{2}Institute of Physics, University of Oldenburg, D-26129 Oldenburg, Germany\\
\authormark{3}Technische Physik, Wilhelm-Conrad-R\"{o}ntgen-Research Center for Complex Material Systems, and W\"{u}rzburg-Dresden Cluster of Excellence ct.qmat, University of W\"{u}rzburg, D-97074 W\"{u}rzburg, Germany}

\email{\authormark{*}maciej.pieczarka@pwr.edu.pl} 



\begin{abstract*}Optical trapping has been proven to be an effective method of separating exciton-polariton condensates from the incoherent high-energy excitonic reservoir located at the pumping laser position. This technique has significantly improved the coherent properties of exciton-polariton condensates, when compared to a quasi-homogeneous spot excitation scheme. Here, we compare two experimental methods on a sample, where a single spot excitation experiment allowed only to observe photonic lasing in the weak coupling regime. In contrast, the ring-shaped excitation resulted in the two-threshold behavior, where an exciton-polariton condensate manifests itself at the first and photon lasing at the second threshold. Both lasing regimes are  trapped in an optical potential created by the pump. We interpret the origin of this confining potential in terms of repulsive interactions of polaritons with the reservoir at the first threshold and as a result of the excessive free-carrier induced refractive index change of the microcavity at the second threshold. This observation offers a way to achieve multiple phases of photonic condensates in samples, e.g., containing novel materials as an active layer, where two-threshold behavior is impossible to achieve with a single excitation spot.
\end{abstract*}

\section{Introduction}

Macroscopic bosonic condensation phenomenon has been one of the central topics at the cross boundary of condensed matter physics and nonlinear optics in recent years \cite{proukakis_snoke_littlewood_2017,Carusotto2013,Wu2019}. The photonic Bose-Einstein condensate (BEC) has been observed in plasmonic lattices \cite{Hakala2018,DeGiorgi2018} and optical microcavities operating either in strong coupling (SC) or weak coupling (WC) regimes \cite{Klaers2010, Balili2007}. Among all these systems, the WC microcavities are a platform where one can observe a BEC of photons or a nonequilibrium photon lasing, depending on the efficiency of thermalization of light by spontaneous emission and absorption in the active medium \cite{Kirton2015}.

In contrast, the SC semiconductor microcavities allow to achieve more phases of photonic BECs, depending on the pumping power strength, i.e., the density of particles. At lower densities, the first threshold is expected at which the system operates in SC regime and the investigated bosonic condensate consists of exciton polaritons - quasiparticles being a coherent superposition of quantum well (QW) excitons and microcavity photons \cite{Kasprzak2006,Balili2007,Carusotto2013}. In this regime, the energy relaxation is efficient due to the interactions between the polaritons and their condensation is enabled by the stimulated scattering to the ground state. At higher powers, the density of particles is large enough to cross the Mott density and to enter the WC regime, where the macroscopic coherence is mediated by the gain medium of the QW electron-hole plasma. In other words, above the second threshold, the sample is expected to work as a standard vertical-cavity surface emitting laser (VCSEL) \cite{Bernard-Duraffourg1961}. Depending on the specific parameters of the sample, it can be characterized by a thermalized Bose-Einstein distribution \cite{Kammann2012,Bajoni2007} or show a nonequilibrium distribution without a well-defined temperature\cite{Kirton2015}.

It is important to highlight that this standard and simplified interpretation of polariton condensation to photon lasing crossover has been questioned in recent years, as more theoretical proposals appeared, predicting a Bardeen-Cooper-Schrieffer (BCS) state \cite{Yamaguchi2013,Byrnes2010,Keeling2005,Kamide2010}, or a dissipative phase transition \cite{Khurgin2020,Hanai2019}. Moreover, the experimental efforts have brought a variety of scenarios, where single, double, and triple thresholds were observed \cite{Schmutzler2013,Sawicki2019,Balili2009,Tempel2012,Bajoni2007,Bajoni2008,Tsotsis2012,Polimeno2020}, as well as signatures of BCS-like polariton lasing. \cite{Hu2021,Horikiri2016}. 

The experimental studies probing the high-density excitation regime share a similar approach of utilizing a quasi-homogeneous spot of the off-resonant pumping laser. Naturally, the creation of a large condensation and lasing area in the sample is advantageous for simpler theoretical descriptions. Nevertheless, condensed particles coexist in this approach with a large density of incoherent high-energy particles, like free electrons, holes, or excitons. It has a detrimental impact on the coherence of the condensed state \cite{Orfanakis2021,Schmutzler2014,Love2008}, changes locally the temperature of the sample \cite{Ballarini2019} or even prevents the observation of the polariton condensate due to the saturation of Rabi splitting \cite{Bajoni2007}. Therefore a method to separate spatially the reservoir from the condensate was required to minimize this influence. One of the first approaches was to create a stress trap \cite{Balili2009,Nelsen2009}. This scheme enabled to locate the excitation laser outside of the trap and accumulate polaritons in the stress-generated potential minimum \cite{Balili2007}. Recently, a more flexible method was developed, namely an optical trapping scheme \cite{Orfanakis2021,Askitopoulos2013,Pieczarka2020,Estrecho2019} in which the condensate is spatially separated from the high-energy incoherent density by shaping the pumping spot. This scheme utilizes the repulsive interactions between polaritons and high-energy reservoir particles, whereby the reservoir forms a complex non-Hermitian potential, providing confinement and gain at the position of the laser pump on the sample \cite{Sun2018,Cristofolini2013,Topfer2020}. To date, the studies on the crossover from polariton condensation to photon lasing remain unexplored in the optical trapping geometry. Here we compare the two geometries of excitation in a SC microcavity and find that a single spot excitation enables the observation of a single threshold and photon lasing, whereas the ring excitation results in a two-threshold behavior and polariton to photon lasing crossover in the confined geometry.

\section{Experimental details}

The sample under study is a high-quality GaAs-based $\lambda/2$ microcavity grown with molecular beam epitaxy. It consist of two distributed Bragg reflectors (DBRs) made of 27 (bottom) and 23.5 (top)  Al$_{0.20}$Ga$_{0.80}$As/AlAs pairs and an active region of 12 GaAs/AlAs QWs of nominal thickness of 13 nm, separated by 4 nm AlAs barriers. The normal mode splitting between the cavity mode and the heavy hole exciton is measured to be about $\hbar\Omega \approx 8$~meV, visible in the single particle dispersion measured at low excitation density in Fig.~1 (a). The experiment was performed on the position of the sample corresponding to the negative detuning $\Delta = E_{Cav} - E_X = - 4.98$~meV.

We utilized a micro-photoluminescence setup with Fourier space imaging \cite{Pieczarka2019}. The real space or Fourier space was imaged on a 0.3~m focal-length monochromator, equipped with a liquid nitrogen cooled CCD camera. The excitation was provided by a single-mode continuous-wave (CW) Ti:sapphire laser, where the beam was chopped with an acousto-optic modulator (AOM) to a train of 5 $\mu$s pulses with a duty cycle of 5\% to prevent excessive heating of the sample. The laser wavelength (energy) was tuned to a high-energy Bragg minimum of the microcavity at around 1.62 eV.

To realize the two different excitation schemes, an additional pair of axicon lenses with an imaging lens was placed in the laser path, similar to the method of Ref.~\cite{Askitopoulos2013}. This effects in a ring-shaped excitation on the sample, as shown in Fig.~1 (b). We used an excitation ring with $20~\mu$m diameter and thickness (full width at half maximum - FWHM) of $3~\mu$m. To obtain a Gaussian spot of similar size, the excitation setup was replaced by a single lens, which defocused the laser spot, as shown in Fig.~1 (c). The obtained excitation spot had a diameter (FWHM) of about $18.2~\mu$m. 

\begin{figure}[ht]
\centering\includegraphics[width=\textwidth]{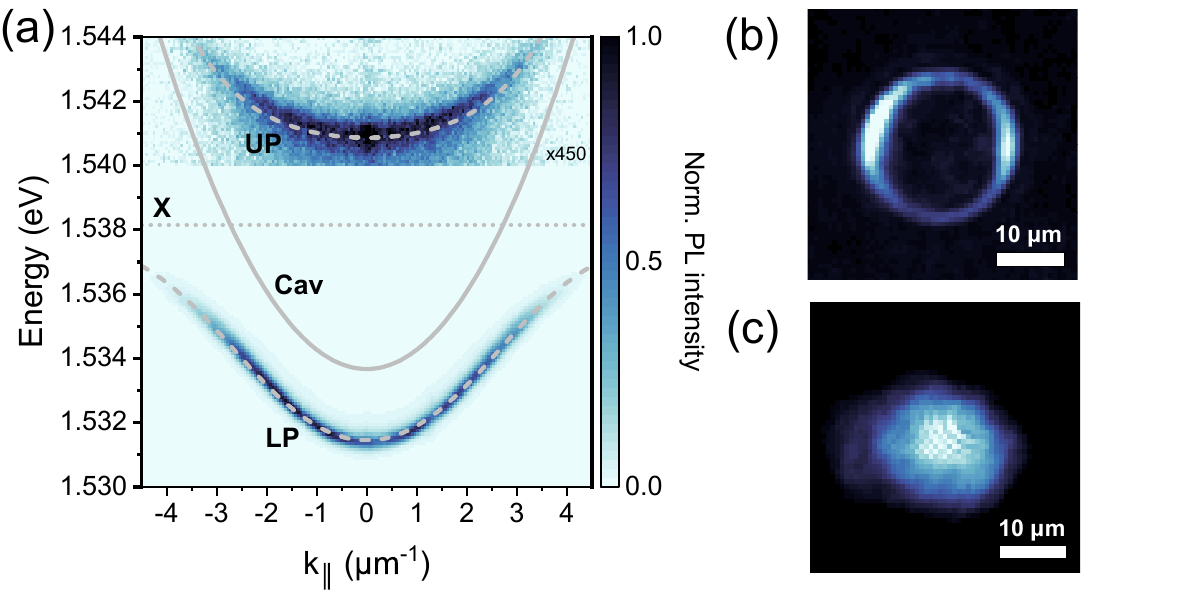}
\caption{(a) Dispersion of polariton branches measured at low excitation density. Theoretical fits with the coupled oscillator model are shown, where upper and lower polariton (UP, LP) branches fit with dashed , bare cavity mode (Cav) with solid and exciton (X) with dotted lines. The UP signal is enhanced 450 times for better visibility. Images of the ring excitation (b) and Gaussian spot (c) reflected from the sample. White bar corresponds to 10~$\mu$m.}
\end{figure}

\section{Results}

We performed power-dependent measurements in both excitation schemes and the result of the standard input-output analysis of the $k_\parallel = 0$ emission is presented in Fig.~2 (the signal is integrated for $|k_\parallel|<0.05~\mu\text{m}^{-1}$). Firstly, let us discuss the Gaussian spot excitation output. A single-threshold behavior is clearly seen in Fig.~2(a), where a nonlinear rise of the intensity happens at $P^{spot}_{th} \approx 15$~mW (power density 5.8 kW/cm$^2$). It is accompanied by a rapid linewidth narrowing, Fig.~2(e), and a blueshift of the emitting mode, coinciding with the energy of the bare cavity mode at the threshold value, as seen in Fig.~2(c). These experimental signatures are typical for a transition to photon lasing in the weak coupling regime\cite{Deng2003,Tsotsis2012,Butte2002}. This interpretation is further supported by inspecting the dispersion of the mode at the threshold power, which is presented in Fig.~3(a). The observed dispersion matches perfectly the expected bare cavity dispersion obtained from the low-density fitting of the single-particle dispersions, as shown in Fig.~1(a). The emission energy of the cavity lasing is further blueshifted at larger excitation powers, which is explained further.

\begin{figure}[ht]
\centering\includegraphics[width=\textwidth]{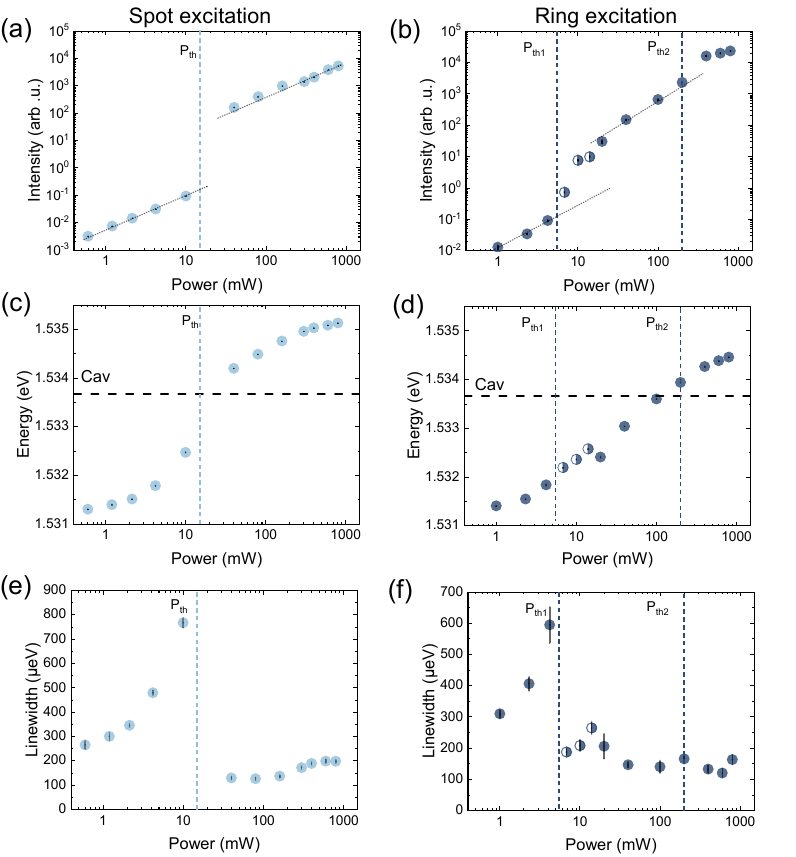}
\caption{Power-dependent measurements of the intensity (a), energy (c) and linewidth (e) at the $k_\parallel =0$ for the spot excitation experiment. The corresponding data for the ring excitation experiment is shown in (b), (d) and (f). "Cav" indicates the bare cavity energy in the sample. The half-filled points for this series indicate an excited state in the trap and full points the ground state. Error bars within the 95\% confidence interval arise from data fitting.} 
\end{figure}

Excitation with the ring-shaped excitation laser beam resulted in an entirely different observation. The input-output curve is now characterized by a distinct two-threshold behavior, where $P_{th1}^{ring} \approx 5.5$~mW (power density 2.9 kW/cm$^2$) and $P_{th2}^{ring} \approx 200$~mW (power density 106 kW/cm$^2$), as shown in Fig~2(b). Polariton condensation is observed at the first threshold, where a massive occupation of excited states confined in the optical trap is observed, see Fig.~3(b), together with linewidth narrowing, shown in Fig.~2(f). This is a common observation for confined exciton-polariton condensates at negative detunings $\Delta<0$, where the photon fraction of the quasiparticle $|C|^2>0.5$, which means a more photonic character for polaritons. The preferential occupation of excited states is a consequence of inefficient relaxation towards the ground state, \cite{Estrecho2019,Pieczarka2020} because of lowered polariton-polariton interactions, and non-Hermitian mode selectivity of the gain distributed at the ring \cite{Sun2018,Askitopoulos2015}. At stronger excitation power, we observed the blueshift of the multimode spectrum and, finally, obtained a polariton density large enough to amplify the energy relaxation process and achieve the largest occupation of the ground state\cite{Estrecho2019} at around $P\approx 20~$mW, see Fig.~3(c). Additionally, within the power range between $P_{th1}^{ring}$ and $P_{th2}^{ring}$, the blueshifted emission energy is located spectrally below the cavity mode, which supports the interpretation of observed polariton condensate is indeed in the SC regime. 

\begin{figure}[ht]
\centering\includegraphics[width=.9\textwidth]{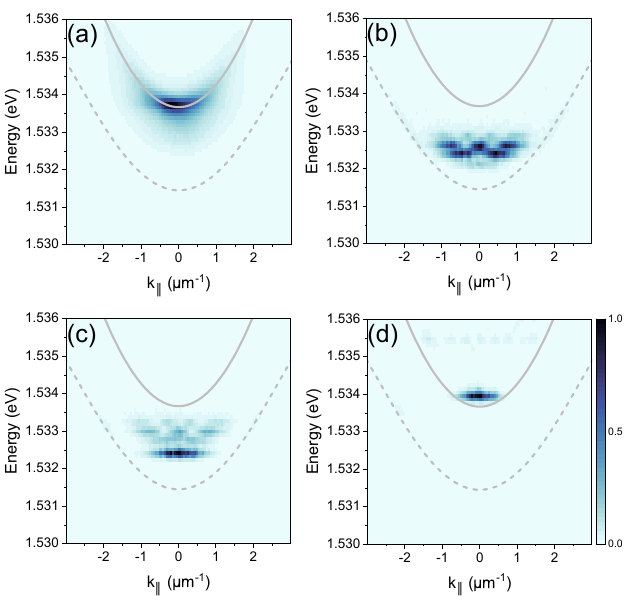}
\caption{(a) Dispersion of the WC lasing at $P_{th}^{spot}$ in the spot excitation experiment. Dispersions in the ring excitation experiment, presenting the excited state polariton condensation at $P=14.4$~mW in (b), ground state relaxation at $P=20$~mW in (c) and WC lasing in (d) at $P_{th2}^{ring}$. }
\end{figure}

Continued increase of the pumping power yields a larger blueshift of the ground state which eventually reaches the energy of the bare cavity mode, Fig.~2(d) and Fig.~3(d). This marks the transition to the WC lasing, as it is accompanied by another sharp increase in the emission at the second threshold at $P_{th2}^{ring}$. Furthermore, we observed the blueshifted ground state emission from the bare cavity mode at $P_{th2}^{ring}$, which we interpret as a confinement effect in the ring geometry. It is confirmed by analyzing the real space spectrum slices through the center of the ring trap above $P_{th1}^{ring}$ for polaritons and $P_{th2}^{ring}$ for photons, as shown in Figs.~4(a) and 4(b). One observes quantized spectra, where the trap ground state as well as the whole excited spectrum in the WC regime is blueshifted from the energy of the polariton ground state in the planar sample at this exciton-photon detuning. Comparison of the pumping power densities in the geometry of the ring excitation with the WC lasing threshold obtained with the Gaussian spot indicates that the WC regime is obtained first at the position of the ring, at the pumping power around $P = 11~\text{mW} \approx P_{th}^{spot}$, at lower powers than the confined WC lasing in the trap. It is a consequence of the higher density of free carriers around the pumped area compared to the center of the trap. However, in ring-shaped excitation the presence of a reservoir was also observed in the trap \cite{Pieczarka2019,Estrecho2021}, therefore, it is the main cause of reaching WC lasing at the highest powers in our experiment. This observation underlines the importance of using the confined geometry that separates most excitonic and free carrier reservoirs from polaritons to achieve polariton condensation before reaching the WC regime in our sample.

The interpretation of optically induced potentials in polariton condensates in SC cavities is well established based on the polariton condensate interaction with the high-energy incoherent reservoir\cite{Askitopoulos2013,Pieczarka2020,Cristofolini2013}. On the other hand, the confinement of the lasing mode in WC microcavities has a different origin. The confinement of light is typically achieved when a cavity has a modified refractive index landscape, which breaks the continuous two-dimensional symmetry of the sample. Changes in the refractive index of the cavity can be induced by several mechanisms. The most common effects are temperature\cite{Anguiano2019}, Kerr-like or free carrier-induced changes of the refractive index\cite{Peinke2021,Pieczarka2020APL,Yuce2012,Xie2016}. Owing to the fact that we observed a continuous blueshift of the lasing mode in the WC regime in both pumping schemes (Fig.~2(c) and Fig.~2(d)), we infer that the dominant source of the refractive index change arises from the excessive density of free carriers, as other effects would result in the opposite sign of the energy shift.

\begin{figure}[ht]
\centering\includegraphics[width=0.75\textwidth]{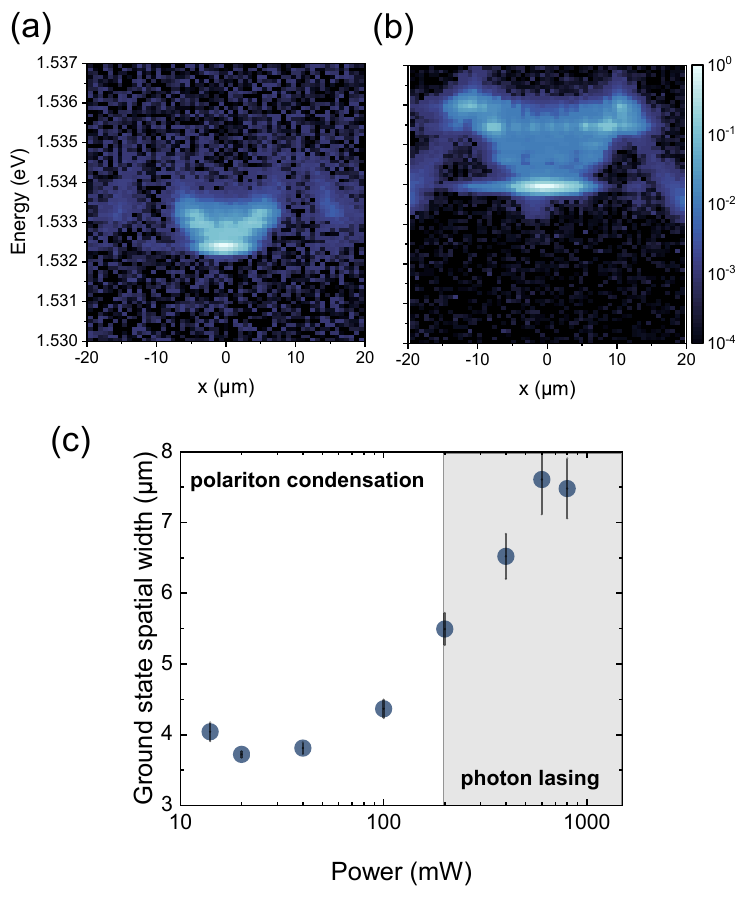}
\caption{Real-space spectra measured at the central slice of the trap for the ground state polariton condensation in (a) and for WC lasing (b). Note that the color scale is normalized and logarithmic. (c) The spatial width of the ground state as a function of the pumping power. Error bars within the 95\% confidence interval arise from data fitting.}
\end{figure}

Surprisingly, the optical potential due to this effect is large and provides a confinement of the optical modes. It is estimated to be at least as high as 2.5 meV (measured from the ground state to the topmost confined mode energy difference in the trap in Fig.~4(b)). We can estimate the change in refractive index from the extracted value of energy shift. The energy of the unperturbed cavity mode is given by $E_{0} = \frac{\hbar c}{nL_{c}}$, where $L_c$ is the cavity length and $n_0$ is the effective refractive index. The change in refractive index locally shifts the mode energy $E_{1} = \frac{\hbar c}{(n+\Delta n)L_{c}}$ and the relative energy change can be approximated by $\frac{\Delta E}{E_{0}}=-\frac{\Delta n}{n} \frac{1}{1+\Delta n/n} \approx -\frac{\Delta n}{n}$, where $\Delta E = E_1-E_0$. In our experiment, the photon confinement potential is caused by a refractive index change of $\frac{\Delta n}{n} \approx -0.16 \%$, which is a value comparable to other experiments \cite{Yuce2012,Xie2016,Harding2007}. This estimation is also confirmed by a similar shift of the lasing mode in our experiment with a Gaussian spot at the same pump power density (visible above the threshold in Fig. 2(c)). Additionally, one could argue that the refractive index change effect also exists in the SC regime; however, it is expected to be negligible since the power and particle densities are much lower than in the WC regime in the experiment.

Another interesting difference between the SC and WC regimes in circular confinement is the actual size of the trap. This can be measured by tracking the spatial size of the ground state as a function of power, as presented in Fig.~4(c). Remarkably, the ground state size is roughly constant for the polariton condensate (about 4~$\mu$m), whereas it grows rapidly at the transition to photon lasing, reaching a spatial size twice as large at the highest pumping powers. This can be interpreted as an effect of spatial hole burning, as the system enters the WC regime \cite{Scot1993}. Importantly, in our sample, spatial hole burning is not observable in the SC\cite{Estrecho2019,Pieczarka2020}. It is because of a relatively low value of the Rabi splitting. At larger pumping powers, the sample easily enters the WC regime, before building up a large enough polariton density, in contrast to the typical high-density BECs in the Thomas-Fermi regime obtained in semiconductor samples with ultra-long photon  lifetimes\cite{Estrecho2019,Pieczarka2020}.

Finally, we compare the emission spectra in the two experiments at the photon lasing threshold. The real space spectra measured at $P_{th2}^{ring}$ and $P_{th}^{spot}$ are presented in Fig.~5. One observes a drastic difference between the two, where the ring excitation yields a sharp lasing peak of the trap ground state with a large occupation of the excited states. The distribution is highly nonequilibrium and no temperature can be associated with this state. Unlike the trapped case, in the spot excitation regime we observed a thermalized spectrum, which shows an exponential decay of occupation of the excited states that can be fit by a Boltzmann distribution. The fit yields a temperature $T=27.9$~K, which is larger than the temperature of the cryostat at 5~K. Thermalized photon distribution with temperature larger than the sample was previously observed at low temperatures, where the thermalization is not complete \cite{Bajoni2007, Kammann2012}. We hypothesize that it might also reflect that the nonresonant excitation heats the sample. Finally, the discrete spectrum in the trapped case shows enhanced occupation of the excited states.  The power density at $P_{th2}^{ring}$ is much larger than for $P_{th}^{spot}$, so we created a larger gain at the ring position in comparison to the Gaussian spot. Additionally, the distribution of the gain on the ring promotes the multimode lasing commonly observed in broad-area VCSELs.

\begin{figure}[ht]
\centering\includegraphics[width=.6\textwidth]{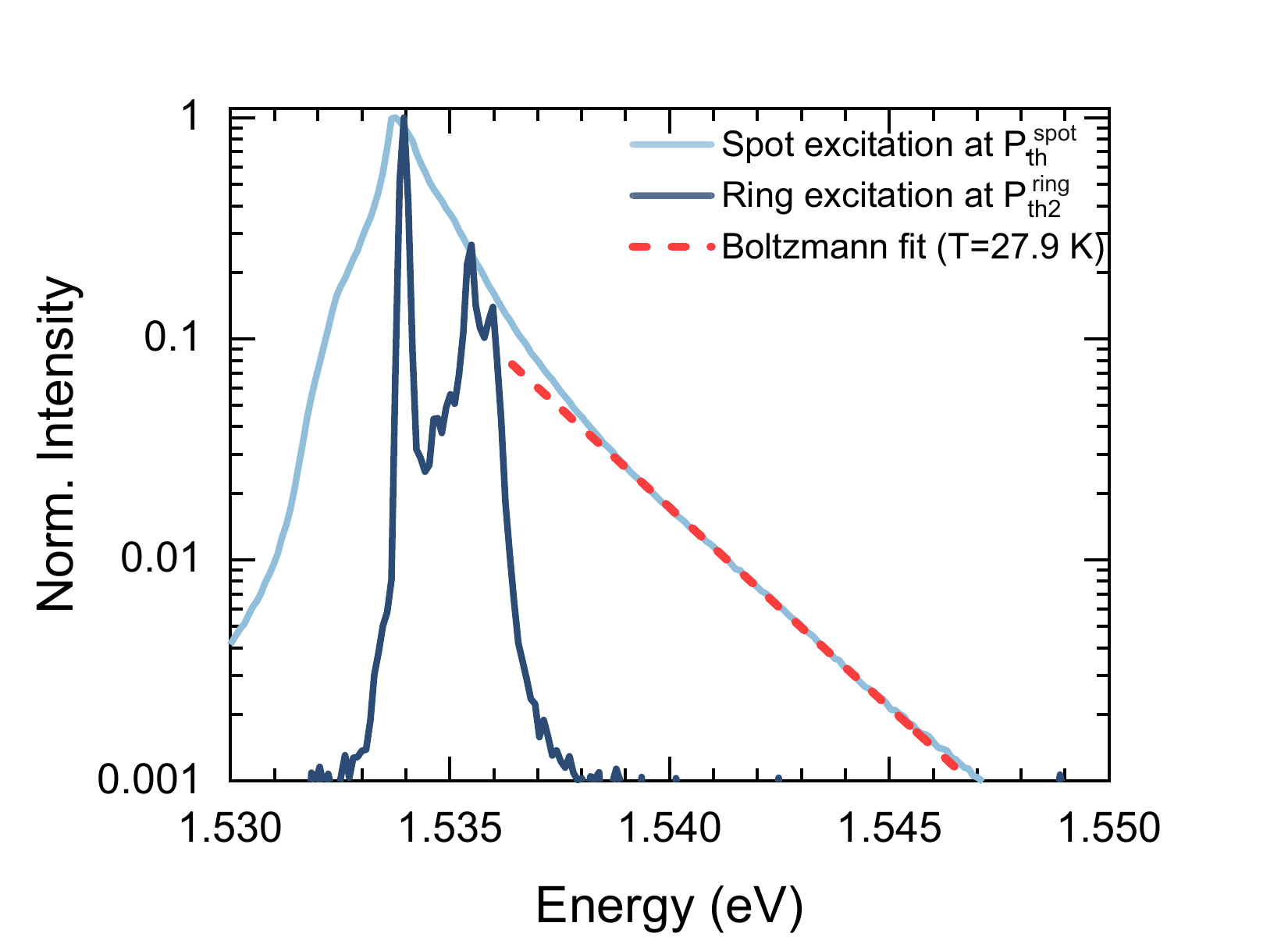}
\caption{Real-space emission spectra at the thresholds corresponding to WC lasing under the spot and ring excitation. The Boltzmann fit to the excited states tail of the spot excitation is plotted with red dashed line.}
\end{figure}

\section{Conclusion}

We have presented a comprehensive experimental comparison of power-dependent studies of a SC microcavity under nonrensonant excitation using a large Gaussian spot and a ring-shaped excitation profile. Under the spot excitation, we observed a transition from SC to WC regime and a single-threshold behavior in power-dependent studies. In the trapped geometry, a two-threshold behavior and crossover from polariton condensation to photon lasing was observed with a distinct spatial hole-burning effect in the photon lasing power region. Additionally, we compared the spectra at photon lasing thresholds and discussed the photon thermalization or lack of thermalization in both approaches. This shows that the two excitation regimes differ significantly and one cannot make definitive conclusions based only on a single approach. Our observations offer an alternative pathway when the saturation of Rabi splitting prevents polariton condensation. Spatial separation of excitation from the condensed state can be used in studying the fundamental properties of particle interactions in the nonlinear regime with reduced influence of incoherent reservoir\cite{Estrecho2019,Pieczarka2020}, yet to be done in polariton condensates in microcavities containing novel materials\cite{Anton-Solanas2021,Yagafarov2020,Betzold2020}. Lastly, the trapped geometry can be used in studying the details of polariton to photon condensation crossover in samples with moderate Rabi splitting, providing a platform to observe dissipative phase transitions \cite{Hanai2019}.

\begin{backmatter}
\bmsection{Funding}
\bmsection{Acknowledgments} 
Wroc\l{}aw group acknowledges support from the Polish National Science Center, Grants No. 2018/30/E/ST7/00648 and No. 2020/39/D/ST3/03546 and from the  the Polish National Agency for Academic Exchange. The W\"{u}rzburg group gratefully acknowledges funding by the State of Bavaria.

\bmsection{Disclosures}
The authors declare no conflicts of interest.

\bmsection{Data availability} Data underlying the results presented in this paper are not publicly available at this time but may be obtained from the authors upon reasonable request.

\end{backmatter}
\bibliography{Bibliography}

\begin{thebibliography}{10}
\newcommand{\enquote}[1]{``#1''}

\bibitem{proukakis_snoke_littlewood_2017}
N.~P. Proukakis, D.~W. Snoke, and P.~B. Littlewood, eds., \emph{Universal
  Themes of Bose-Einstein Condensation} (Cambridge University, 2017).

\bibitem{Carusotto2013}
I.~Carusotto and C.~Ciuti, \enquote{{Quantum fluids of light},}
  {\protect\JournalTitle{Reviews of Modern Physics}} \textbf{85}, 299--366
  (2013).

\bibitem{Wu2019}
F.~O. Wu, A.~U. Hassan, and D.~N. Christodoulides, \enquote{{Thermodynamic
  theory of highly multimoded nonlinear optical systems},}
  {\protect\JournalTitle{Nature Photonics}} \textbf{13}, 776--782 (2019).

\bibitem{Hakala2018}
T.~K. Hakala, A.~J. Moilanen, A.~I. V{\"{a}}kev{\"{a}}inen, R.~Guo, J.~P.
  Martikainen, K.~S. Daskalakis, H.~T. Rekola, A.~Julku, and
  P.~T{\"{o}}rm{\"{a}}, \enquote{{Bose-Einstein condensation in a plasmonic
  lattice},} {\protect\JournalTitle{Nature Physics}} \textbf{14}, 739--744
  (2018).

\bibitem{DeGiorgi2018}
M.~{De Giorgi}, M.~Ramezani, F.~Todisco, A.~Halpin, D.~Caputo, A.~Fieramosca,
  J.~Gomez-Rivas, and D.~Sanvitto, \enquote{{Interaction and Coherence of a
  Plasmon-Exciton Polariton Condensate},} {\protect\JournalTitle{ACS
  Photonics}} \textbf{5}, 3666--3672 (2018).

\bibitem{Klaers2010}
J.~Klaers, J.~Schmitt, F.~Vewinger, and M.~Weitz, \enquote{{Bose-Einstein
  condensation of photons in an optical microcavity.}}
  {\protect\JournalTitle{Nature}} \textbf{468}, 545--548 (2010).

\bibitem{Balili2007}
R.~Balili, V.~Hartwell, D.~Snoke, L.~Pfeiffer, and K.~West,
  \enquote{{Bose-Einstein condensation of microcavity polaritons in a trap},}
  {\protect\JournalTitle{Science}} \textbf{316}, 1007--1010 (2007).

\bibitem{Kirton2015}
P.~Kirton and J.~Keeling, \enquote{{Thermalization and breakdown of
  thermalization in photon condensates},} {\protect\JournalTitle{Physical
  Review A}} \textbf{91}, 033826 (2015).

\bibitem{Kasprzak2006}
J.~Kasprzak, M.~Richard, S.~Kundermann, A.~Baas, P.~Jeambrun, J.~M. Keeling,
  F.~M. Marchetti, M.~H. Szym{\'{a}}nska, R.~Andr{\'{e}}, J.~L. Staehli,
  V.~Savona, P.~B. Littlewood, B.~Deveaud, and L.~S. Dang,
  \enquote{{Bose-Einstein condensation of exciton polaritons},}
  {\protect\JournalTitle{Nature}} \textbf{443}, 409--414 (2006).

\bibitem{Bernard-Duraffourg1961}
M.~G.~A. Bernard and G.~Duraffourg, \enquote{Laser conditions in
  semiconductors,} {\protect\JournalTitle{Phys. Status Solidi B}} \textbf{1},
  699--703 (1961).

\bibitem{Kammann2012}
E.~Kammann, H.~Ohadi, M.~Maragkou, A.~V. Kavokin, and P.~G. Lagoudakis,
  \enquote{{Crossover from photon to exciton-polariton lasing},}
  {\protect\JournalTitle{New Journal of Physics}} \textbf{14}, 105003 (2012).

\bibitem{Bajoni2007}
D.~Bajoni, P.~Senellart, A.~Lema{\^{i}}tre, and J.~Bloch, \enquote{{Photon
  lasing in GaAs microcavity: Similarities with a polariton condensate},}
  {\protect\JournalTitle{Physical Review B}} \textbf{76}, 201305 (2007).

\bibitem{Yamaguchi2013}
M.~Yamaguchi, K.~Kamide, R.~Nii, T.~Ogawa, and Y.~Yamamoto, \enquote{{Second
  thresholds in BEC-BCS-laser crossover of exciton-polariton systems},}
  {\protect\JournalTitle{Physical Review Letters}} \textbf{111}, 026404 (2013).

\bibitem{Byrnes2010}
T.~Byrnes, T.~Horikiri, N.~Ishida, and Y.~Yamamoto, \enquote{{BCS wave-function
  approach to the BEC-BCS crossover of exciton-polariton condensates},}
  {\protect\JournalTitle{Physical Review Letters}} \textbf{105}, 186402 (2010).

\bibitem{Keeling2005}
J.~Keeling, P.~R. Eastham, M.~H. Szymanska, and P.~B. Littlewood,
  \enquote{{BCS-BEC crossover in a system of microcavity polaritons},}
  {\protect\JournalTitle{Physical Review B}} \textbf{72}, 115320 (2005).

\bibitem{Kamide2010}
K.~Kamide and T.~Ogawa, \enquote{{What determines the wave function of
  electron-hole Pairs in polariton condensates?}}
  {\protect\JournalTitle{Physical Review Letters}} \textbf{105}, 056401 (2010).

\bibitem{Khurgin2020}
J.~B. Khurgin, \enquote{{Exceptional points in polaritonic cavities and
  subthreshold Fabry–Perot lasers},} {\protect\JournalTitle{Optica}}
  \textbf{7}, 1015 (2020).

\bibitem{Hanai2019}
R.~Hanai, A.~Edelman, Y.~Ohashi, and P.~B. Littlewood, \enquote{{Non-Hermitian
  Phase Transition from a Polariton Bose-Einstein Condensate to a Photon
  Laser},} {\protect\JournalTitle{Phys. Rev. Lett.}} \textbf{122}, 185301
  (2019).

\bibitem{Schmutzler2013}
J.~Schmutzler, F.~Veit, M.~A{\ss}mann, J.~S. Tempel, S.~H{\"{o}}fling, M.~Kamp,
  A.~Forchel, and M.~Bayer, \enquote{{Determination of operating parameters for
  a GaAs-based polariton laser},} {\protect\JournalTitle{Appl. Phys. Lett.}}
  \textbf{102}, 081115 (2013).

\bibitem{Sawicki2019}
K.~Sawicki, J.~G. Rousset, R.~Rudniewski, W.~Pacuski, M.~{\'{S}}ciesiek,
  T.~Kazimierczuk, K.~Sobczak, J.~Borysiuk, M.~Nawrocki, and
  J.~Suffczy{\'{n}}ski, \enquote{{Triple threshold lasing from a photonic trap
  in a Te/Se-based optical microcavity},} {\protect\JournalTitle{Communications
  Physics}} \textbf{2}, 38 (2019).

\bibitem{Balili2009}
R.~Balili, B.~Nelsen, D.~W. Snoke, L.~Pfeiffer, and K.~West, \enquote{Role of
  the stress trap in the polariton quasiequilibrium condensation in gaas
  microcavities,} {\protect\JournalTitle{Phys. Rev. B}} \textbf{79}, 075319
  (2009).

\bibitem{Tempel2012}
J.~S. Tempel, F.~Veit, M.~A{\ss}mann, L.~E. Kreilkamp, A.~Rahimi-Iman,
  A.~L{\"{o}}ffler, S.~H{\"{o}}fling, S.~Reitzenstein, L.~Worschech,
  A.~Forchel, and M.~Bayer, \enquote{{Characterization of two-threshold
  behavior of the emission from a GaAs microcavity},}
  {\protect\JournalTitle{Phys. Rev. B}} \textbf{85}, 075318 (2012).

\bibitem{Bajoni2008}
D.~Bajoni, P.~Senellart, E.~Wertz, I.~Sagnes, A.~Miard, A.~Lema\^{\i}tre, and
  J.~Bloch, \enquote{Polariton laser using single micropillar
  $\mathrm{GaAs}\mathrm{\text{\ensuremath{-}}}\mathrm{GaAlAs}$ semiconductor
  cavities,} {\protect\JournalTitle{Phys. Rev. Lett.}} \textbf{100}, 047401
  (2008).

\bibitem{Tsotsis2012}
P.~Tsotsis, P.~S. Eldridge, T.~Gao, S.~I. Tsintzos, Z.~Hatzopoulos, and P.~G.
  Savvidis, \enquote{{Lasing threshold doubling at the crossover from strong to
  weak coupling regime in GaAs microcavity},} {\protect\JournalTitle{New
  Journal of Physics}} \textbf{14}, 023060 (2012).

\bibitem{Polimeno2020}
L.~Polimeno, A.~Fieramosca, G.~Lerario, M.~Cinquino, M.~{De Giorgi},
  D.~Ballarini, F.~Todisco, L.~Dominici, V.~Ardizzone, M.~Pugliese, C.~T.
  Prontera, V.~Maiorano, G.~Gigli, L.~{De Marco}, and D.~Sanvitto,
  \enquote{{Observation of Two Thresholds Leading to Polariton Condensation in
  2D Hybrid Perovskites},} {\protect\JournalTitle{Advanced Optical Materials}}
  \textbf{8}, 2000176 (2020).

\bibitem{Hu2021}
J.~Hu, Z.~Wang, S.~Kim, H.~Deng, S.~Brodbeck, C.~Schneider, S.~H{\"{o}}fling,
  N.~H. Kwong, and R.~Binder, \enquote{{Polariton Laser in the
  Bardeen-Cooper-Schrieffer Regime},} {\protect\JournalTitle{Physical Review
  X}} \textbf{11}, 011018 (2021).

\bibitem{Horikiri2016}
T.~Horikiri, M.~Yamaguchi, K.~Kamide, Y.~Matsuo, T.~Byrnes, N.~Ishida,
  A.~L{\"{o}}ffler, S.~H{\"{o}}fling, Y.~Shikano, T.~Ogawa, A.~Forchel, and
  Y.~Yamamoto, \enquote{{High-energy side-peak emission of exciton-polariton
  condensates in high density regime},} {\protect\JournalTitle{Sci. Rep.}}
  \textbf{6}, 25655 (2016).

\bibitem{Orfanakis2021}
K.~Orfanakis, A.~F. Tzortzakakis, D.~Petrosyan, P.~G. Savvidis, and H.~Ohadi,
  \enquote{{Ultralong temporal coherence in optically trapped exciton-polariton
  condensates},} {\protect\JournalTitle{Physical Review B}} \textbf{103},
  235313 (2021).

\bibitem{Schmutzler2014}
J.~Schmutzler, T.~Kazimierczuk, {\"{O}}.~Bayraktar, M.~A{\ss}mann, M.~Bayer,
  S.~Brodbeck, M.~Kamp, C.~Schneider, and S.~H{\"{o}}fling, \enquote{{Influence
  of interactions with noncondensed particles on the coherence of a
  one-dimensional polariton condensate},} {\protect\JournalTitle{Physical
  Review B}} \textbf{89}, 115119 (2014).

\bibitem{Love2008}
A.~P.~D. Love, D.~N. Krizhanovskii, D.~M. Whittaker, R.~Bouchekioua,
  D.~Sanvitto, S.~A. Rizeiqi, R.~Bradley, M.~S. Skolnick, P.~R. Eastham,
  R.~Andr\'e, and L.~S. Dang, \enquote{Intrinsic decoherence mechanisms in the
  microcavity polariton condensate,} {\protect\JournalTitle{Physical Review
  Letters}} \textbf{101}, 067404 (2008).

\bibitem{Ballarini2019}
D.~Ballarini, I.~Chestnov, D.~Caputo, M.~De~Giorgi, L.~Dominici, K.~West, L.~N.
  Pfeiffer, G.~Gigli, A.~Kavokin, and D.~Sanvitto, \enquote{Self-trapping of
  exciton-polariton condensates in gaas microcavities,}
  {\protect\JournalTitle{Physical Review Letters}} \textbf{123}, 047401 (2019).

\bibitem{Nelsen2009}
B.~Nelsen, R.~Balili, D.~W. Snoke, L.~Pfeiffer, and K.~West, \enquote{Lasing
  and polariton condensation: Two distinct transitions in gaas microcavities
  with stress traps,} {\protect\JournalTitle{J. Appl.Phys.}} \textbf{105},
  122414 (2009).

\bibitem{Askitopoulos2013}
A.~Askitopoulos, H.~Ohadi, A.~V. Kavokin, Z.~Hatzopoulos, P.~G. Savvidis, and
  P.~G. Lagoudakis, \enquote{{Polariton condensation in an optically induced
  two-dimensional potential},} {\protect\JournalTitle{Physical Review B}}
  \textbf{88}, 041308 (2013).

\bibitem{Pieczarka2020}
M.~Pieczarka, E.~Estrecho, M.~Boozarjmehr, O.~Bleu, M.~Steger, K.~West, L.~N.
  Pfeiffer, D.~W. Snoke, J.~Levinsen, M.~M. Parish, A.~G. Truscott, and E.~A.
  Ostrovskaya, \enquote{{Observation of quantum depletion in a non-equilibrium
  exciton–polariton condensate},} {\protect\JournalTitle{Nature
  Communications}} \textbf{11}, 429 (2020).

\bibitem{Estrecho2019}
E.~Estrecho, T.~Gao, N.~Bobrovska, D.~Comber-Todd, M.~D. Fraser, M.~Steger,
  K.~West, L.~N. Pfeiffer, J.~Levinsen, M.~M. Parish, T.~C.~H. Liew,
  M.~Matuszewski, D.~W. Snoke, A.~G. Truscott, and E.~A. Ostrovskaya,
  \enquote{{Direct measurement of polariton-polariton interaction strength in
  the Thomas-Fermi regime of exciton-polariton condensation},}
  {\protect\JournalTitle{Physical Review B}} \textbf{100}, 035306 (2019).

\bibitem{Sun2018}
Y.~Sun, Y.~Yoon, S.~Khan, L.~Ge, M.~Steger, L.~N. Pfeiffer, K.~West, H.~E.
  T{\"{u}}reci, D.~W. Snoke, and K.~A. Nelson, \enquote{{Stable switching among
  high-order modes in polariton condensates},} {\protect\JournalTitle{Physical
  Review B}} \textbf{97}, 045303 (2018).

\bibitem{Cristofolini2013}
P.~Cristofolini, A.~Dreismann, G.~Christmann, G.~Franchetti, N.~G. Berloff,
  P.~Tsotsis, Z.~Hatzopoulos, P.~G. Savvidis, and J.~J. Baumberg,
  \enquote{Optical superfluid phase transitions and trapping of polariton
  condensates,} {\protect\JournalTitle{Phys. Rev. Lett.}} \textbf{110}, 186403
  (2013).

\bibitem{Topfer2020}
J.~D. T\"opfer, H.~Sigurdsson, S.~Alyatkin, and P.~G. Lagoudakis,
  \enquote{Lotka-volterra population dynamics in coherent and tunable
  oscillators of trapped polariton condensates,} {\protect\JournalTitle{Phys.
  Rev. B}} \textbf{102}, 195428 (2020).

\bibitem{Pieczarka2019}
M.~Pieczarka, M.~Boozarjmehr, E.~Estrecho, Y.~Yoon, M.~Steger, K.~West, L.~N.
  Pfeiffer, K.~A. Nelson, D.~W. Snoke, A.~G. Truscott, and E.~A. Ostrovskaya,
  \enquote{{Effect of optically induced potential on the energy of trapped
  exciton-polaritons below the condensation threshold},}
  {\protect\JournalTitle{Physical Review B}} \textbf{100}, 85301 (2019).

\bibitem{Deng2003}
H.~Deng, G.~Weihs, D.~Snoke, J.~Bloch, and Y.~Yamamoto, \enquote{Polariton
  lasing vs. photon lasing in a semiconductor microcavity,}
  {\protect\JournalTitle{Proceedings of the National Academy of Sciences}}
  \textbf{100}, 15318--15323 (2003).

\bibitem{Butte2002}
R.~Butt\'e, G.~Delalleau, A.~I. Tartakovskii, M.~S. Skolnick, V.~N. Astratov,
  J.~J. Baumberg, G.~Malpuech, A.~Di~Carlo, A.~V. Kavokin, and J.~S. Roberts,
  \enquote{Transition from strong to weak coupling and the onset of lasing in
  semiconductor microcavities,} {\protect\JournalTitle{Phys. Rev. B}}
  \textbf{65}, 205310 (2002).

\bibitem{Askitopoulos2015}
A.~Askitopoulos, T.~C.~H. Liew, H.~Ohadi, Z.~Hatzopoulos, P.~G. Savvidis, and
  P.~G. Lagoudakis, \enquote{Robust platform for engineering pure-quantum-state
  transitions in polariton condensates,} {\protect\JournalTitle{Phys. Rev. B}}
  \textbf{92}, 035305 (2015).

\bibitem{Estrecho2021}
E.~Estrecho, M.~Pieczarka, M.~Wurdack, M.~Steger, K.~West, L.~N. Pfeiffer,
  D.~W. Snoke, A.~G. Truscott, and E.~A. Ostrovskaya, \enquote{{Low-Energy
  Collective Oscillations and Bogoliubov Sound in an Exciton-Polariton
  Condensate},} {\protect\JournalTitle{Phys. Rev. Lett.}} \textbf{126}, 75301
  (2021).

\bibitem{Anguiano2019}
S.~Anguiano, A.~A. Reynoso, A.~E. Bruchhausen, A.~Lema{\^{i}}tre, J.~Bloch, and
  A.~Fainstein, \enquote{{Three-dimensional trapping of light with light in
  semiconductor planar microcavities},} {\protect\JournalTitle{Physical Review
  B}} \textbf{99}, 195308 (2019).

\bibitem{Peinke2021}
E.~Peinke, T.~Sattler, G.~M. Torelly, P.~L. Souza, S.~Perret, J.~Bleuse,
  J.~Claudon, W.~L. Vos, and J.-M. G{\'{e}}rard, \enquote{{Tailoring the
  properties of quantum dot-micropillars by ultrafast optical injection of free
  charge carriers},} {\protect\JournalTitle{Light: Science \& Applications}}
  \textbf{10}, 215 (2021).

\bibitem{Pieczarka2020APL}
M.~Pieczarka, D.~Poletti, C.~Schneider, S.~H{\"{o}}fling, E.~A. Ostrovskaya,
  G.~S{\c{e}}k, and M.~Syperek, \enquote{{Observation of gain-pinned
  dissipative solitons in a microcavity laser},} {\protect\JournalTitle{APL
  Photonics}} \textbf{5}, 086103 (2020).

\bibitem{Yuce2012}
E.~Y{\"{u}}ce, G.~Ctistis, J.~Claudon, E.~Dupuy, K.~J. Boller, J.-M.
  G{\'{e}}rard, and W.~L. Vos, \enquote{{Competition between electronic Kerr
  and free-carrier effects in an ultimate-fast optically switched semiconductor
  microcavity},} {\protect\JournalTitle{Journal of the Optical Society of
  America B}} \textbf{29}, 2630 (2012).

\bibitem{Xie2016}
W.~Xie, F.-K. Hsu, Y.-S. Lee, S.-D. Lin, and C.~W. Lai,
  \enquote{{Multiple-pulse lasing from an optically induced harmonic
  confinement in a highly photoexcited microcavity},}
  {\protect\JournalTitle{Optica}} \textbf{3}, 1477 (2016).

\bibitem{Harding2007}
P.~J. Harding, T.~G. Euser, Y.~R. Nowicki-Bringuier, J.~M. Ǵrard, and W.~L.
  Vos, \enquote{{Dynamical ultrafast all-optical switching of planar GaAsAlAs
  photonic microcavities},} {\protect\JournalTitle{Applied Physics Letters}}
  \textbf{91}, 111103 (2007).

\bibitem{Scot1993}
J.~Scott, R.~Geels, S.~Corzine, and L.~Coldren, \enquote{Modeling temperature
  effects and spatial hole burning to optimize vertical-cavity surface-emitting
  laser performance,} {\protect\JournalTitle{IEEE Journal of Quantum
  Electronics}} \textbf{29}, 1295--1308 (1993).

\bibitem{Anton-Solanas2021}
C.~Anton-Solanas, M.~Waldherr, M.~Klaas, H.~Suchomel, T.~H. Harder, H.~Cai,
  E.~Sedov, S.~Klembt, A.~V. Kavokin, S.~Tongay, K.~Watanabe, T.~Taniguchi,
  S.~H{\"{o}}fling, and C.~Schneider, \enquote{{Bosonic condensation of
  exciton–polaritons in an atomically thin crystal},}
  {\protect\JournalTitle{Nature Materials}} \textbf{20}, 1233--1239 (2021).

\bibitem{Yagafarov2020}
T.~Yagafarov, D.~Sannikov, A.~Zasedatelev, K.~Georgiou, A.~Baranikov,
  O.~Kyriienko, I.~Shelykh, L.~Gai, Z.~Shen, D.~Lidzey, and P.~Lagoudakis,
  \enquote{{Mechanisms of blueshifts in organic polariton condensates},}
  {\protect\JournalTitle{Communications Physics}} \textbf{3}, 18 (2020).

\bibitem{Betzold2020}
S.~Betzold, M.~Dusel, O.~Kyriienko, C.~P. Dietrich, S.~Klembt, J.~Ohmer,
  U.~Fischer, I.~A. Shelykh, C.~Schneider, and S.~Hoefling, \enquote{{Coherence
  and Interaction in Confined Room-Temperature Polariton Condensates with
  Frenkel Excitons},} {\protect\JournalTitle{ACS Photonics}} \textbf{7}, 384
  (2020).

\end{thebibliography}

\end{document}